# PROTEIN SECONDARY STRUCTURE PREDICTION USING DEEP CONVOLUTIONAL NEURAL FIELDS

Sheng Wang*,1,2, Jian Peng[3], Jianzhu Ma[1], and Jinbo Xu*,1

[1] Toyota Technological Institute at Chicago, Chicago, IL
[2] Department of Human Genetics, University of Chicago, Chicago, IL
[3] Department of Computer Science, University of Illinois at Urbana-Champaign, Urbana, IL
* To whom correspondence should be addressed to:
  wangsheng@uchicago.edu, jinboxu@gmail.com

ABSTRACT

Protein secondary structure (SS) prediction is important for studying protein structure and function. When only the sequence (profile) information is used as input feature, currently the best predictors can obtain ~80% Q3 accuracy, which has not been improved in the past decade. Here we present DeepCNF (Deep Convolutional Neural Fields) for protein SS prediction. DeepCNF is a Deep Learning extension of Conditional Neural Fields (CNF), which is an integration of Conditional Random Fields (CRF) and shallow neural networks. DeepCNF can model not only complex sequence-structure relationship by a deep hierarchical architecture, but also interdependency between adjacent SS labels, so it is much more powerful than CNF. Experimental results show that DeepCNF can obtain ~84% Q3 accuracy, ~85% SOV score, and ~72% Q8 accuracy, respectively, on the CASP and CAMEO test proteins, greatly outperforming currently popular predictors. As a general framework, DeepCNF can be used to predict other protein structure properties such as contact number, disorder regions, and solvent accessibility.

INTRODUCTION

The 3D structure of a protein is determined largely by its amino acid sequence[1]. However it is extremely challenging to predict protein structure from sequence alone[2]. Since protein structure is critical to analysis of its function and many applications like drug and/or enzyme design[3-5], understanding the complex sequence-structure relationship is one of the greatest challenges in computational biology[6-8]. Accurate protein structure and function prediction relies, in part, on the accuracy of secondary structure prediction[9-12].

Protein secondary structure (SS) refers to the local conformation of the polypeptide backbone of proteins. There are two regular SS states: alpha-helix (H) and beta-strand (E), as suggested by Pauling[13] more than 60 years ago, and one irregular SS type: coil region (C). Sander[14] developed a DSSP algorithm to classify SS into 8 fine-grained states. In particular, DSSP assigns 3 types for helix (G for $3_{10}$ helix, H for alpha-helix, and I for pi-helix), 2 types for strand (E for beta-strand and B for beta-bridge), and 3 types for coil (T for beta-turn, S for high curvature loop, and L for irregular). Overall, protein secondary structure can be regarded as a bridge that links the primary sequence and tertiary structure and thus, is used by many structure and functional analysis tools[15-18].

Protein SS prediction has been extensively studied[10-12,19-35]. Many computational methods have been developed to predict both 3-state SS and a few to predict 8-state SS. Meanwhile, 8-state prediction may provide more detailed local structure information[33,34,36]. Holley & Karplus[19] and Qian & Sejnowski[20] may be the first that used neural networks (NN) to predict SS, which have been followed by a few others[19,21,23,24,37]. The most significant improvement in SS prediction was achieved by Rost & Sander[23] and Zvelebil *et. al*[35] by making use of sequence profile derived from multiple sequence alignment[38-40]. Jones *et.al.*[24] developed a 2-stage neural network method PSIPRED, which takes PSI-BLAST sequence profile[41] as input and obtains ~80% accuracy for 3-state SS prediction. Other machine learning methods include bidirectional recurrent neural networks[26,34,37] (which can capture spatial dependency), probabilistic graphical models[25,29,42], support vector machines [27,28,30,43] and hidden Markov models[22,31].

Very recently Baldi *et.al.*[34] presented a template-based method for SS prediction, which can yield much better accuracy by making use of solved structures as templates. However, when close templates are not available, Baldi's method performs slightly worse than PSIPRED. Cheng *et.al.*[44] proposed a deep learning approach to 3-state SS prediction using a typical deep belief network model, in which each layer is a restricted Boltzmann machine (RBM)[45] and trained by contrastive divergence[46] in an unsupervised manner. Zhou & Troyanskaya[36] reported another deep learning approach to 8-state SS prediction using a supervised generative stochastic network, which to our best knowledge may be the best 8-state predictor. However, neither Cheng nor Zhou reported a better than 80% accuracy for 3-state prediction.

SS prediction is usually evaluated by Q3 or Q8 accuracy, which measures the percent of residues for which 3-state or 8-state secondary structure is correctly predicted[44]. So far the best Q3 accuracy for ab initio prediction (i.e., templates are not allowed) is ~80% obtained by PSIPRED and a few other state-of-the-art approaches such as JPRED[47,48]. It is very challenging to develop a method that can break this long-lasting record. This may be because the relatively shallow architectures of existing methods cannot model well the complex sequence-structure relationship. Alternatively, 3-state SS prediction could also be measured by segment of overlap (SOV) score, which can be interpreted as SS segment-based accuracy. SOV allows for small wrong predictions at SS segment ends, but penalizes more on wrong predictions in the middle region of a SS segment[49].

In this paper we present a machine learning method DeepCNF (Deep Convolutional Neural Fields) for both 3-state and 8-state SS prediction. DeepCNF combines the advantages of both conditional neural fields (CNF)[50] and deep convolutional neural networks (DCNN)[51], which captures not only complex sequence-structure relationship, but also models SS label correlation among adjacent residues. DeepCNF is similar to conditional random fields (CRF)[52] and CNF[33] in modeling interdependency among adjacent SS labels. However, DeepCNF uses DCNN to replace the shallow neural networks used in CNF so that it can capture very complex relationship between input features and output labels. This DCNN can also include longer-range sequence information (see Figure 1 and 2).

Our DeepCNF method differs from Cheng's method[44] in that the latter uses a typical deep belief network (see Figure 1A) while we use a deep convolutional network (see Figure 1B). As such, our method can capture longer-range sequence information than Cheng's method. Our method also differs from Cheng's method in that the latter does not explicitly model SS interdependency among adjacent residues. Our method differs from Zhou's deep learning method (denoted as ICML2014)[36] in the following aspects: (1) our method places only input features in a visible layer and treats the SS labels as hidden states while Zhou's method places both the input features and SS labels in a visible layer; (2) our method explicitly models the SS label interdependency while Zhou's method does not; (3) our method directly calculates the conditional probability of SS labels on input features while Zhou's method uses sampling; (4) our method trains the model parameter simultaneously from the input to output layer while Zhou's method trains the model parameters layer-by-layer; and (5) more importantly, our method demonstrated a significantly improved Q3 accuracy and SOV score while Zhou's method did not.

Our experiments show that our method greatly outperforms the state-of-the-art methods, especially on those structure types which are more challenging to predict, such as high curvature regions (S), beta loop (T), and irregular loop (L).

RESULTS

**Dataset.** We used five publicly available datasets: (1) CullPDB[53] of 6125 proteins, (2) CB513 of 513 proteins, (3) CASP10[54] and (4) CASP11[55] datasets containing 123 and 105 domain sequences, respectively, and (5) CAMEO (http://www.cameo3d.org/sp/6-months/) test proteins in the past 6 months (from 2014-12-05 to 2015-05-29). Meanwhile, datasets (2-5) are only used for test. The CullPDB dataset was constructed before CASP10 (i.e., May 2012) and any two proteins in this set share less than 25% sequence identity with each other. Following the same procedure in[36], we divided CullPDB into two subsets for training and test, respectively, such that the training proteins share no more than 25% sequence identity with our test sets (2-4). Our training set consists of ~5600 CullPDB proteins and the remaining ~500 PDB proteins are used as the test data. In total there are 403 CAMEO test targets in the past 6 months and 179 proteins are kept for test after removing those sharing more than 25% sequence identity with the training set. The native SS labels of all the training and test proteins are generated by DSSP[14].

An alternative way to select non-redundant proteins for training and test is to pick one representative from each protein superfamily defined by CATH[56] or SCOP[57]. By using test proteins in different superfamilies than the training proteins, we can reduce the bias incurred by the sequence profile similarity between the training and test proteins. To fulfill this, we use the publically available JPRED training and test data[47] (http://www.compbio.dundee.ac.uk/jpred4/about.shtml), which has 1338 training and 149 test proteins, respectively, each of which belongs to a different superfamily.

**Programs to compare.** We compare our method DeepCNF-SS (abbreviated as DeepCNF) with the following programs: SSpro[34], RaptorX-SS8[33], ICML2014[36] for 8-state SS prediction; and SSpro, RaptorX-SS8, PSIPRED[24], SPINE-X[12], JPRED[47], for 3-state SS prediction. The SSpro package uses two prediction strategies: without template and with template (i.e., using a solved structure in PDB as template). All the other test methods do not make use of template information at all. All the programs are run with their parameters set according to their respective papers. The program derived from the ICML2014 method is not publicly available, so we cannot evaluate its performance on CASP10, CASP11 and CAMEO test sets. We cannot test Cheng's deep learning method either because it is not publicly available. However, only minor improvement in Q3 accuracy over PSIPRED was reported by Cheng's paper[44].

**Performance metric.** We measure the prediction results in terms of Q3 and Q8 accuracy. The Q3 (Q8) accuracy is defined as the percentage of residues for which the predicted secondary structures are correct[32]. For 3-state SS prediction, we also calculate the SOV (Segment of OVerlap) score[49], which measures how well the observed and the predicted SS segments match. In particular, the SOV measure assigns a lower score to the prediction deviating from observed SS segment length distribution even if it has high Q3 accuracy (i.e., per-residue accuracy)[31]. A wrong prediction in the middle region of a SS segment results in a lower SOV score than a wrong prediction at the terminal regions. A detailed definition of SOV is described in[32], and also in our Supplemental File.

**Determining the regularization factor by cross validation.** Our DeepCNF has only a hyper-parameter, i.e., the regularization factor, which is used to avoid overfitting. Once it is fixed, we can estimate all the model parameters by solving the optimization problem in Eq. (10). To choose the right regularization factor and examine the stability of our DeepCNF model, we conduct a five-fold cross-validation test. In particular, we randomly divide the training set (containing 5600 CullPDB proteins) into 5 subsets and then use 4 subsets as training and one as validation. Figure 3 shows the Q8 accuracy of our DeepCNF model with respect to the regularization factor. The optimal regularization factor is around 50, which yields 73.2% Q8 accuracy on average. At this point, the Q8 accuracy difference of all the models is less than 1%, consistent with the previous report[33].

**Determining the DeepCNF architecture.** The architecture of our DeepCNF model is mainly determined by the following 3 factors (see Figure 2): (i) the number of hidden layers; (ii) the number of different neurons at each layer; and (iii) the window size at each layer. We fix the window size to 11 because the average length of an alpha helix is around eleven residues[58] and that of a beta strand is around six[59]. To show the relationship between the performance and the number of hidden layers, we trained four different DeepCNF models with 1, 3, 5, and 7 layers, respectively. All the models use the same window size (i.e., 11) and the same number (i.e., 100) of different neurons at each layer. In total these four models have ~50K, ~270K, ~500K, and ~700K parameters, respectively. We trained the models with different regularization factors. As shown in Figure 4A, when only one hidden layer is used, DeepCNF becomes CNF[50] and its performance is quite similar to RaptorX-SS8 (single model) as shown in Table 1. When more layers are applied, the Q8 accuracy gradually improves. To balance the model complexity and performance, by default we set window size to 11 and use 5 hidden layers, each with 100 different neurons.

To show that it is the deep convolutional structure but not the number of model parameters that mainly contributes to performance improvement, we trained four models with ~500K model parameters but 4 different numbers of layers: 1, 3, 5 and 7. We still use the same window size at 11 and for each model all the layers have the same number of neurons. That is, for the 1-, 3-, 5- and 7-layer models, we use 1000, 140, 100, and 85 neurons for each layer, respectively. As shown in Figure 4B, the models with more layers have better Q8 accuracy, although they have the same number of model parameters. Since the 7-layer model is only slightly better than the 5-layer model, to reduce computational complexity, in the following experimental results we use a DeepCNF model of 5 hidden layers and 100 neurons for each layer. The window size at each layer is set to 11.

**Overall performances.** Tables 1, 2 and 3 show the Q8, Q3 accuracy, and SOV score of our method DeepCNF and the others on the five datasets. As listed in these tables, when templates are not used, our method significantly outperforms the others, including the popular PSIPRED, our old method RaptorX-SS8[33], SPINE-X and the recent deep learning method ICML2014[36]. SSpro with template obtains very good accuracy on CullPDB, CB513 and CASP10, but not on CASP11 and CAMEO. This is because SSpro uses a template database built in 2013, covering only the former three sets but not CASP11 or CAMEO. Most CASP11 and CAMEO test proteins share <25% sequence identity with any template databases created before 2014. In terms of Q3 accuracy on CASP10, DeepCNF obtains 84.4%, even slightly outperforming SSpro with template (84.2%). In terms of both Q3 and Q8 accuracy on CASP11, DeepCNF obtains 83.8% and 71.9%, respectively, significantly outperforming SSpro with template (78.4% and 66.7%, respectively). The same trend is also observed on 179 CAMEO targets, where DeepCNF obtains 84.4% Q3 and 72.1% Q8 accuracy, respectively, much better than SSpro with template (78.9% and 65.7%, respectively).

When only CASP10 and CASP11 hard targets (number is 85) are evaluated, DeepCNF, PSIPRED, SPINE-X, JPRED, RaptorX-SS8, SSpro (with template) and SSpro (without template) have Q3 accuracy 82.2%, 77.9%, 77.1%, 78.5%, 76.2%, 76.9 and 75.4%, respectively, and SOV score 83.0%, 78.1%, 77.2%, 79.8%, 77.9%, 75.3% and 73.6%, respectively. All the methods have lower Q3 accuracy on these hard targets than on the whole datasets since all the test methods use sequence profiles as input features and a hard target usually has sparse sequence profile that carries little evolutionary information. In addition, the hard targets do not have good structure homologs in the training set, so a predictor cannot copy SS labels from the training data as prediction. However, our method even has a slightly larger advantage over the others on the CASP hard targets than on the whole CASP sets. This implies that our method is slightly better than the others in dealing with sparse sequence profiles and learning sequence-structure relationship from the training data. Similar results are observed on the 86 CAMEO hard targets, on which DeepCNF, PSIPRED, SPINE-X, JPRED, RaptorX-SS8, SSpro (with template) and SSpro (without template) have Q3 accuracy 82.1%, 78.0%, 77.6%, 77.7%, 76.8%, 77.0%, and 76.5%, respectively, and SOV score 81.7%, 77.1%, 74.8%, 78.2%, 73.7% , 72.6%, and 72.5%, respectively.

The SOV score tolerates on small wrong predictions at the terminal regions of a segment while penalizes more on the erroneous predictions in the middle region of a segment[49]. As shown in Table 3, in terms of SOV score on all the five datasets, DeepCNF obtains 86.2%, 84.8%, 85.6%, 85.8%, and 84.5%, respectively, significantly outperforming all the other predictors including SSpro with template. These results show that DeepCNF could yield more meaningful SS predictions. This is mainly because our deep convolutional neural networks are better in predicting beta turn (T), high curvature loop (S), and irregular loop (L) states, which appear more often at the boundary of a helix or sheet segment. The other reason is that the conditional random fields in our method models the interdependency among adjacent residues in a SS segment, which helps reduce erroneous predictions in the middle region of a segment.

**Recall and precision.** Table 4 shows the recall and precision on each of the 8 states obtained by our method DeepCNF and the second best method ICML2014[36] on the CullPDB test set. Both methods fail on state I since it is too rare to even appear in the test set. The result of the ICML 2014 method is taken from its paper. Overall, our method obtains better recall and precision for each state, especially those non-ordinary states such as G, S and T. For the high curvature loop (S), our method has recall and precision 0.323 and 0.543, respectively, while ICML2014 obtains 0.159 and 0.423. For beta turn (T), our method obtains recall and precision 0.594 and 0.613, respectively, while ICML2014 obtains 0.506 and 0.548, respectively. DeepCNF also outperforms ICML2014 for the loop (L) state. This result could be due to the fact that S, T, and L state may be impacted by medium-range information on the protein sequence[36] and our method is better than the others in modeling this kind of information. Our DeepCNF has a similar performance trend on the CB513 test set (see Table 5). DeepCNF outperforms ICML2014, especially on those non-ordinary states, as well as the ordinary beta sheet state. The largest advantage lies in predicting curvature loop (S) and $3_{10}$ helix (G). We cannot do a detailed comparison between our method and the ICML2014 method on the other test sets since the program derived from the ICML2014 method is not publicly available.

**Prediction accuracy with respect to homologous information.** We further examine the performance of DeepCNF with respect to the amount of homologous information measured by Neff[33]. The Neff of a protein measures the average number of effective amino acids across all the residues, ranging from 1 to 20 since there are only 20 amino acids in nature. A small Neff indicates that the protein has a sparse sequence profile. By contrast, a large Neff implies that the protein may have a large amount of homologous information. Figure 5 shows the Q3 accuracy of the five tested methods on the CB513 and the two CASP datasets with respect to Neff. When Neff≤2, DeepCNF performs slightly better than the others. However, when Neff>2, DeepCNF greatly outperforms the others.

**Where is the improvement from?** We used 25% sequence identity as cutoff to remove redundancy between the training and test sets. Although the training and test proteins may not have similar primary sequences, their sequence profiles may be still similar, so one may wonder if our improvement is due to the sequence profile similarity between the test and training proteins. We conducted one stricter experiment to study this problem. In particular, we retrained our DeepCNF models using the 1338 JPRED training proteins and tested the resultant models on the 149 JPRED test proteins[47]. All the test and training proteins belong to different superfamilies. That is, it is unlikely that one test protein shares similar sequence profile with one training protein. The sequence profiles of these JPRED training and test proteins are generated from an NR database dated in 2012-10-26. We divided the training set into 7 groups according to the JPRED cross-validation sets (available at http://www.compbio.dundee.ac.uk/jpred4/about.shtml) and then trained 7 DeepCNF models separately. Each model is trained by the proteins in 6 groups. Tested on the 149 JPRED test proteins, the resultant 7 DeepCNF models have an average Q3 accuracy of 84.9% (see Supplemental Table 1), far better than what can be obtained by the other methods. For example, on this test set, JPRED, which is one of the best SS predictors, has Q3 accuracy 82.1%.

This experimental result indicates that the improvement mainly comes from the DeepCNF model itself instead of the profile similarity between the test and training proteins. In fact, as shown in previous sections, our method also greatly outperforms the others on the CASP11 and CAMEO hard targets (which do not have similar profiles as our training proteins), which further confirms this conclusion.

## CONCLUSION AND FUTURE WORK

We have presented a new sequence labeling method, called DeepCNF (Deep Convolutional Neural Fields), for protein secondary structure prediction. This new method can not only model complex sequence-structure relationship by a deep hierarchical architecture, but also exploit interdependency between adjacent SS labels. The overall performance of DeepCNF is significantly better than the state-of-the-art methods, breaking the long-lasting ~80% Q3 accuracy[34]. DeepCNF is even better than the other methods in terms of SOV score. In particular, DeepCNF performs much better on the SS types which are challenging to predict, such as high curvature region (state S), beta loop (state T), and irregular loop (state L). DeepCNF also performs reasonably well on proteins without any good homologs in PDB, better than the other methods. However, DeepCNF has no significant advantage over the others when a protein under prediction has very sparse sequence profile (i.e., Neff≤2). That is, it is still challenging to predict SS structure from primary sequence instead of sequence profile.

In addition to secondary structure prediction, DeepCNF can be directly applied to many sequence labelling problems[50,60-62]. For example, DeepCNF can be used to predict solvent accessibility[34,63,64], contact number[65], structural alphabet[66-69] and order/disorder regions[70-72], which are useful for other purposes such as protein threading, remote homology detection[73-75], and protein model quality assessment[76,77].

METHOD

## DeepCNF model

As shown in Figure 2, DeepCNF consists of two modules: (a) the Conditional Random Fields (CRF) module consisting of the top layer and the label layer, and (b) the deep convolutional neural network (DCNN) module covering the input to the top layer. When only one hidden layer is used, this DeepCNF becomes Conditional Neural Fields (CNF), a probabilistic graphical model described in[50].

**Conditional Random Field (CRF).** Given a protein sequence of length $L$, let $\boldsymbol{Y} = (Y_1, \ldots, Y_L)$ denote its SS where $Y_i$ is the SS type at residue $i$. Let $\boldsymbol{X} = (X_1, \ldots, X_L)$ denote the input feature where $X_i$ is a column vector representing the input feature for residue $i$. Using DeepCNF, we calculate the conditional probability of $\boldsymbol{Y}$ on the input $\boldsymbol{X}$ as follows,

$$P(\boldsymbol{Y}|\boldsymbol{X}) = \exp(\sum_{i=1}^{L}[\Psi(\boldsymbol{Y},\boldsymbol{X},i) + \Phi(\boldsymbol{Y},\boldsymbol{X},i)])/Z(\boldsymbol{X}) \quad (1)$$

Where $\Psi(\boldsymbol{Y},\boldsymbol{X},i)$ is the potential function quantifying correlation among adjacent SS types at around position $i$, $\Phi(\boldsymbol{Y},\boldsymbol{X},i)$ is the potential function modeling relationship between $Y_i$ and input features for position $i$, and $Z(\boldsymbol{X})$ is the partition function. Formally, $\Psi()$ and $\Phi()$ are defined as follows,

$$\Psi(\boldsymbol{Y},\boldsymbol{X},i) = \sum_{a,b} T_{a,b} \delta(Y_i = a)\delta(Y_{i+1} = b) \quad (2)$$

$$\Phi(\boldsymbol{Y},\boldsymbol{X},i) = \sum_{a} \sum_{m} U_{a,m} H_m(\boldsymbol{X},i,W)\delta(Y_i = a) \quad (3)$$

Where $a$ and $b$ represent secondary structure states, $\delta()$ is an indicator function, $H_m(\boldsymbol{X},i,W)$ is a neural network function for the m-th neuron at position $i$ of the top layer, and $W$, $U$, and $T$ are the model parameters to be trained. Specifically, $W$ is the parameter for the neural network, $U$ is the parameter connecting the top layer to the label layer, and $T$ is for label correlation. Below see the details of the deep convolutional neural network for $H_m(\boldsymbol{X},i,W)$.

**Deep convolutional neural network (DCNN).** Figure 6 shows two adjacent layers. Let $M_k$ be the number of neurons for a single position at the $k$-th layer. Let $X_i(m)$ be the $m$-th feature at the input layer for residue $i$ and $H_i^k(m)$ denote the output value of the $m$-th neuron of position $i$ at layer $k$. When $k = 1$, $\boldsymbol{H}^k$ is actually the input feature $\boldsymbol{X}$. Otherwise, $\boldsymbol{H}^k$ is a matrix of dimension $L \times M_k$. Let $2N_k + 1$ be the window size at the $k$-th layer. Mathematically, $H_i^k(m)$ is defined as follows.

$$H_i^k(m) = X_i(m), \qquad \text{if } k = 1$$

$$H_i^{k+1}(m) = h(\sum_{n=-N_k}^{N_k} \sum_{m'=1}^{M_k} [H_{i+n}^k(m') * W_n^k(m,m')]), \qquad \text{if } k < K$$

$$H_m(\boldsymbol{X},i,W) = H_i^k(m) \qquad \text{if } k = K \quad (4)$$

Meanwhile, $h()$ is the activation function, either the sigmoid (i.e., $1/(1 + \exp(-x))$) or the tanh (i.e., $(1 - \exp(-2x))/(1 + \exp(-2x))$) function. $W_n^k$ where $(-N_k \leq n \leq N_k)$ is a 2D weight matrix for the connections between the neurons of position $i + n$ at layer $k$ and the neurons of position $i$ at layer $k + 1$. $W_n^k(m,m')$ is shared by all the positions in the same layer, so it is position-independent. Here $m'$ and $m$ index two neurons at the $k$-th and $(k + 1)$-th layers, respectively.

## Training method

Similar to CRF[52], we train the model parameters by maximum-likelihood. The log-likelihood is as follows.

$$\log P(\boldsymbol{Y}|\boldsymbol{X}) = \sum_{i=1}^{L}[\Psi(\boldsymbol{Y},\boldsymbol{X},i) + \Phi(\boldsymbol{Y},\boldsymbol{X},i)] - \log Z(\boldsymbol{X}) \quad (5)$$

To train the model parameters, we need to calculate the gradient with respect to each parameter. We calculate the gradient first for CRF and then for DCNN. The gradient of the log-likelihood with respect to the parameters $T$ and $U$ is given by,

$$\nabla_{T_{a,b}} = [\sum_{i=1}^{L} \delta(Y_i = a)\delta(Y_{i+1} = b)] - EXP_{P(\widetilde{Y}|X,W,U,T)}[\sum_{i=1}^{L} \delta(\widetilde{Y}_i = a)\delta(\widetilde{Y}_{i+1} = b)] \quad (6)$$

$$\nabla_{U_{a,m}} = [\sum_{i=1}^{L} \delta(Y_i = a)H_m(X, i, W)] - EXP_{P(\widetilde{Y}|X,W,U,T)}[\sum_{i=1}^{L} \delta(\widetilde{Y}_i = a)H_m(X, i, W)] \quad (7)$$

Where $EXP$ is the expectation function and can be calculated efficiently using the forward-backward algorithm[50,52].

As shown in Figure 7, we can calculate the neuron error values at the $k$-th layer in a back-propagation mode as follows.

$$E_i^k(m) = g(H_i^k(m)) * \sum_a [E_i(a) * U_{a,m}] \quad , \qquad \text{if } k = K$$
$$\text{where} \quad E_i(a) = [\delta(Y_i = a)] - EXP_{P(\widetilde{Y}|X,W,U,T)}[\delta(\widetilde{Y}_i = a)]$$

$$E_i^k(m) = g(H_i^k(m)) * \sum_{n=-N_k}^{N_k} \sum_{m'=1}^{M_{k+1}} [E_{i-n}^{k+1}(m') * W_n^k(m', m)], \qquad \text{if } k < K \quad (8)$$

Where $g()$ is the derivative of the activation function; it is $g(x) = (1 - x)x$ and $g(x) = 1 - x^2$ for the sigmoid and tanh function, respectively. $E^k$ is the neuron error value matrix at the $k$-th layer, with dimension $L \times M_k$. $E_i(a)$ is the error of the log-likelihood function with respect to the label at the $i$-th position and can be calculated by the forward-backward algorithm. Finally, the gradient of the parameter $W$ at the $k$-th layer is:

$$\nabla_{W_n^k(m,m')} = \sum_{i=1}^{L} [E_i^{k+1}(m) * H_{i+n}^k(m')] \quad (9)$$

L2 regularization and L-BFGS

To reduce over-fitting, the log-likelihood objective function is penalized with a $L_2$-norm of the model parameters. Thus, our final objective function is as follows.

$$\max_\theta \log P_\theta(Y|X) - \lambda \|\theta\|^2 \quad (10)$$

Where $\theta$ is the set of model parameters and $\lambda$ is the regularization factor used to avoid overfitting. Although DeepCNF has a large number of model parameters, by setting the regularization factor large enough, we can make the $L_2$-norm of the model parameters small and thus, restrict the search space of the model parameter and avoid overfitting. However, a very large regularization factor (e.g., infinity) may restrict the model parameter into too small a search space and the resultant model may not learn enough from the training data (i.e., under-fitting). We will determine the regularization factor by cross-validation.

Since the log-likelihood function is not convex, usually we can only solve the objective function to a local instead of global optimum. Although a typical way to train a deep network is to do it layer-by-layer, in our implementation we train all the model parameters simultaneously. We use the L-BFGS[78] to search for the optimal model parameters, which has also been successfully used to train CRF and CNF[50,52]. In addition to learning the model parameters simultaneously, we can also train them layer-by-layer in a supervised mode. Starting from the first layer (i.e., input feature), we train the model parameter $W1$ by removing the third to the $K$-th layers but keeping the label layer. After $W1$ is trained, we generate the neuron output values for the second layer and use them as input to train the model parameter $W2$ by removing the fourth to the $K$-th layers but keeping the label layer. We repeat this procedure until all the parameters are trained, and finally we fine-tune these parameters by simultaneous training.

Input Features

We used the input features described in[36]. In particular, for each protein sequence, we ran PSI-BLAST[41] with E-value threshold 0.001 and 3 iterations to search UniRef90[79] to generate the position specific scoring matrix (PSSM). We then transform PSSM by the sigmoid function $1/(1 + \exp(-x))$ where $x$ is a PSSM entry. We also use a binary vector of 21 elements to indicate the amino acid type at position $i$. We use 21 since there might be some unknown amino acids in a protein sequence. In total there are 42 input features for each residue, 21 from PSSM and the other 21 from the primary sequence. Note that besides using PSSM generated by 3 iterations, we could also add the PSSM generated by 5 iterations into the input features.

Availability

DeepCNF is available at http://raptorx.uchicago.edu/download/.

ACKNOWLEDGEMENTS

*Funding*: National Institutes of Health R01GM0897532 to JX and National Science Foundation DBI-0960390 to JX. The authors are also grateful to the computing power provided by the UChicago Beagle and RCC allocations.

AUTHOR CONTRIBUTIONS

SW conceived, designed and implemented the algorithm. JP and JM helped design the algorithm. SW and JX designed the experiments, analyzed the data and wrote the manuscript. All authors read and approved the final manuscript.

COMPETING FINANCIAL INTERESTS

The authors declare no competing financial interests

FIGURE LEGENDS

**Figure 1.** A typical deep neural network (A) vs. a convolutional deep neural network (B). A convolutional deep neural network can capture longer-range sequence information than a typical deep neural network when both use the same window size.

**Figure 2.** The architecture of DeepCNF, where $i$ is the residue index and $X_i$ the associated input features, $H^k$ represents the $k$-th hidden layer, and $Y$ is the output label. All the layers from the 1st to the top layer form a deep convolutional neural network (DCNN) with parameter $W^k\{k = 1,2,…,K\}$. The top layer and the label layer form a conditional random field (CRF) with $U$ and $T$ being the model parameters. $U$ is the parameter used to connect the top layer to the label layer, and $T$ is used to model correlation among adjacent residues.

**Figure 3.** Five-fold cross-validation results of Q8 accuracy on the CullPDB training set with different regularization factors.

**Figure 4.** The Q8 accuracy on CB513 by the models of different number of layers of 1, 3, 5, and 7 (the same window size is used). (A) Each layer of the 4 models has 100 neurons for a position. The total parameter number of the 4 models is different. (B) Each layer of the models has different neurons for a position. The total parameter number of the 4 models is similar.

**Figure 5.** Q3 accuracy on CB513 and two CASP (CASP10-11) test sets with respect to Neff. Each point represents the average Q3 accuracy on those proteins falling into an Neff interval.

**Figure 6.** The feed-forward connection between two adjacent layers in the deep convolutional neural network.

**Figure 7.** Illustration of calculating the gradient of deep convolutional neural network from layer $k + 1$ to layer $k$.

TABLES

**Table 1.** Q8 accuracy of the tested methods on 5 datasets: CullPDB, CB513, CASP10, CASP11 and CAMEO. The program for ICML2014 is not publicly available. Its result is taken from its paper.

| **Methods** | **Q8(%)** | | | | |
|---|---|---|---|---|---|
| | **CullPDB** | **CB513** | **CASP10** | **CASP11** | **CAMEO** |
| SSpro (without template) | 66.6 | 63.5 | 64.9 | 65.6 | 63.5 |
| SSpro (with template) | **85.1** | **89.9** | **75.9** | 66.7 | 65.7 |
| ICML2014 | 72.1 | 66.4 | - | - | |
| RaptorX-SS8 | 69.7 | 64.9 | 64.8 | 65.1 | 66.2 |
| DeepCNF-SS | 75.2 | 68.3 | 71.8 | **72.3** | **72.1** |

**Table 2.** Q3 accuracy of the tested methods on 5 datasets: CullPDB, CB513, CASP10, CASP11 and CAMEO.

| **Methods** | **Q3(%)** | | | | |
|---|---|---|---|---|---|
| | **CullPDB** | **CB513** | **CASP10** | **CASP11** | **CAMEO** |
| SSpro (without template) | 79.5 | 78.5 | 78.5 | 77.6 | 77.5 |
| SSpro (with template) | **88.7** | **90.7** | 84.2 | 78.4 | 78.9 |
| SPINE-X | 81.7 | 78.9 | 80.7 | 79.3 | 80.0 |
| PSIPRED | 82.5 | 79.2 | 81.2 | 80.7 | 80.1 |
| JPRED | 82.9 | 81.7 | 81.6 | 80.4 | 79.7 |
| RaptorX-SS8 | 81.2 | 78.3 | 78.9 | 79.1 | 79.4 |
| DeepCNF-SS | 85.4 | 82.3 | **84.4** | **84.7** | **84.5** |

**Table 3.** SOV score of the tested methods on 5 datasets: CullPDB, CB513, CASP10, CASP11 and CAMEO.

| Methods | SOV score (%) | | | | |
|---|---|---|---|---|---|
| | CullPDB | CB513 | CASP10 | CASP11 | CAMEO |
| SSpro (without template) | 77.4 | 77.2 | 75.9 | 77.3 | 75.4 |
| SSpro (with template) | 81.3 | 79.4 | 80.7 | 77.4 | 76.3 |
| SPINE-X | 79.1 | 78.7 | 78.7 | 79.3 | 79.4 |
| PSIPRED | 81.8 | 81.0 | 80.9 | 81.4 | 80.1 |
| JPRED | 82.5 | 83.3 | 82.4 | 82.0 | 80.7 |
| RaptorX-SS8 | 80.9 | 79.5 | 80.2 | 81.1 | 78.1 |
| DeepCNF-SS | **86.7** | **84.8** | **85.7** | **86.5** | **85.5** |

**Table 4.** Recall and precision of DeepCNF and ICML2014 on the CullPDB test set.

| SS8 label | Recall | | Precision | |
|---|---|---|---|---|
| | DeepCNF | ICML2014 | DeepCNF | ICML2014 |
| L | **0.707** | 0.633 | **0.615** | 0.541 |
| B | **0.046** | 0.001 | **0.638** | 0.5 |
| E | **0.867** | 0.823 | **0.814** | 0.748 |
| G | **0.302** | 0.133 | **0.535** | 0.496 |
| I | **0.0** | 0.0 | **0.0** | 0.0 |
| H | **0.937** | 0.935 | **0.878** | 0.828 |
| S | **0.323** | 0.159 | **0.543** | 0.423 |
| T | **0.594** | 0.506 | **0.613** | 0.548 |

**Table 5.** Recall and precision of DeepCNF and ICML2014 on the CB513 dataset.

| SS8 label | Recall | | Precision | |
|---|---|---|---|---|
| | DeepCNF | ICML2014 | DeepCNF | ICML2014 |
| L | **0.657** | 0.655 | **0.571** | 0.518 |
| B | **0.026** | 0.0 | **0.433** | 0.0 |
| E | **0.833** | 0.797 | **0.748** | 0.717 |
| G | **0.26** | 0.131 | **0.49** | 0.45 |
| I | **0.0** | 0.0 | **0.0** | 0.0 |
| H | **0.904** | 0.9 | **0.849** | 0.831 |
| S | **0.255** | 0.14 | **0.487** | 0.444 |
| T | **0.528** | 0.503 | **0.53** | 0.496 |

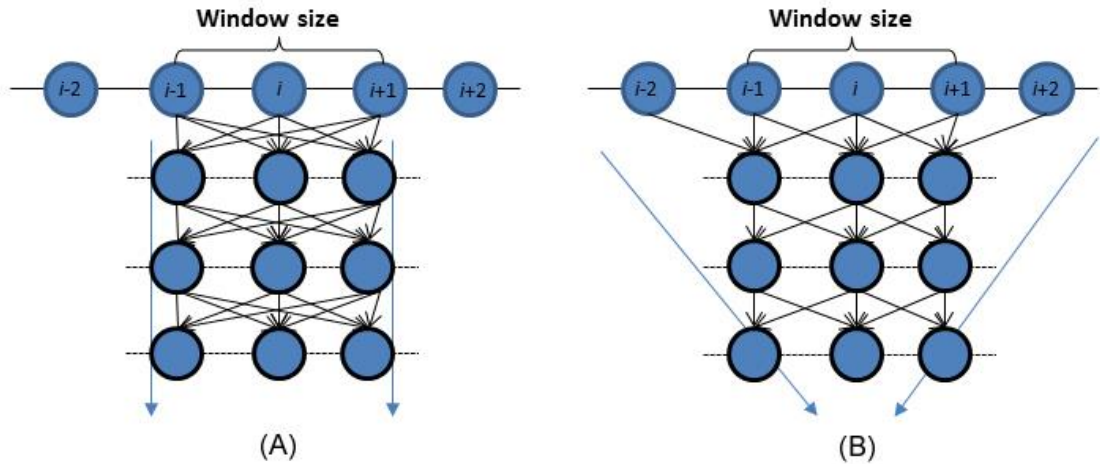

**Figure 1.** A typical deep neural network (A) vs. a convolutional deep neural network (B). A convolutional deep neural network can capture longer-range sequence information than a typical deep neural network when both use the same window size.

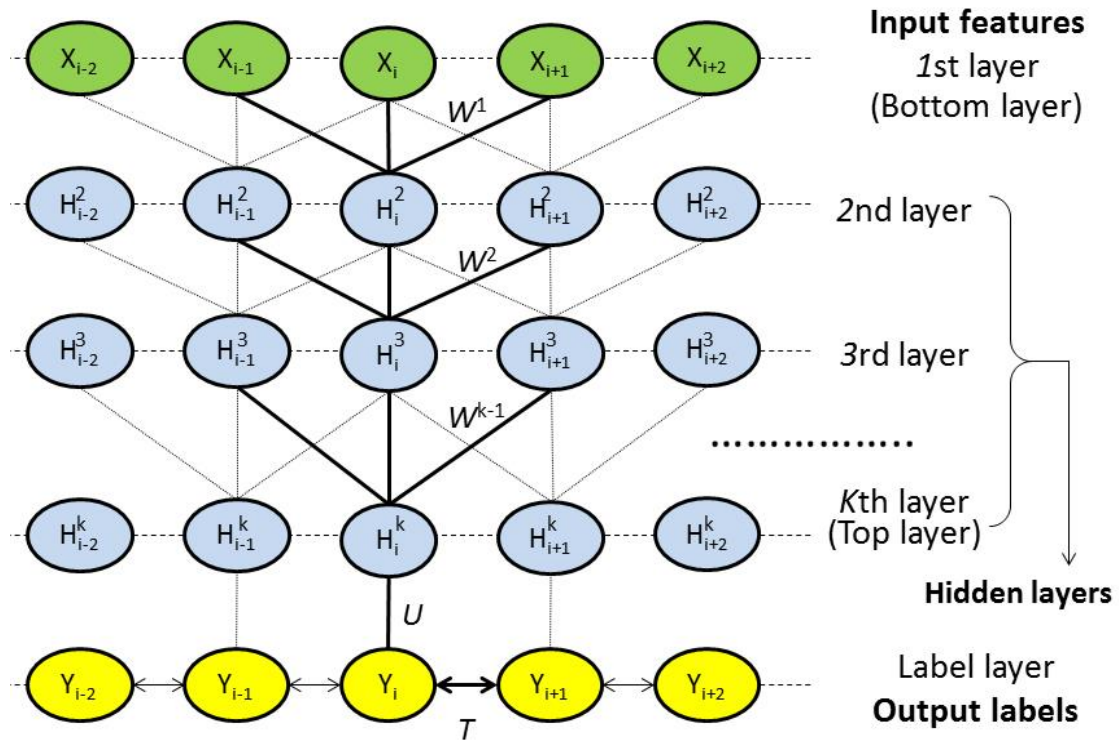

**Figure 2.** The architecture of DeepCNF, where $i$ is the residue index and $X_i$ the associated input features, $H^k$ represents the $k$-th hidden layer, and $Y$ is the output label. All the layers from the 1st to the top layer form a deep convolutional neural network (DCNN) with parameter $W^k\{k = 1,2,\ldots,K\}$. The top layer and the label layer form a conditional random field (CRF) with $U$ and $T$ being the model parameters. $U$ is the parameter used to connect the top layer to the label layer, and $T$ is used to model correlation among adjacent residues.

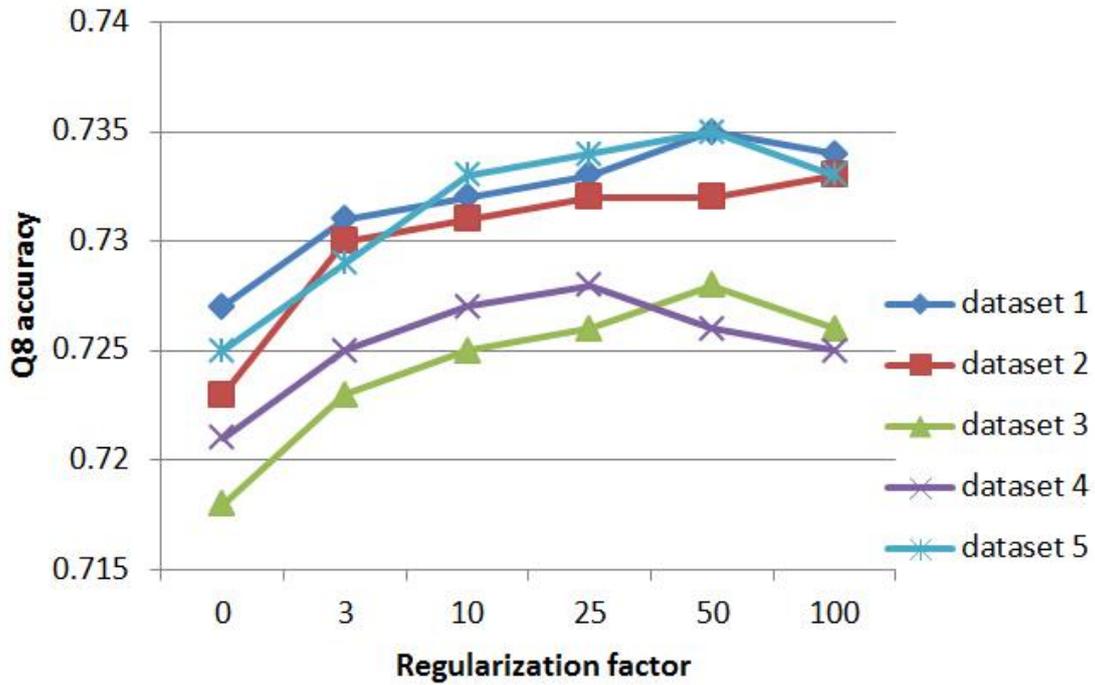

**Figure 3.** Five-fold cross-validation results of Q8 accuracy on the CullPDB training set with different regularization factors.

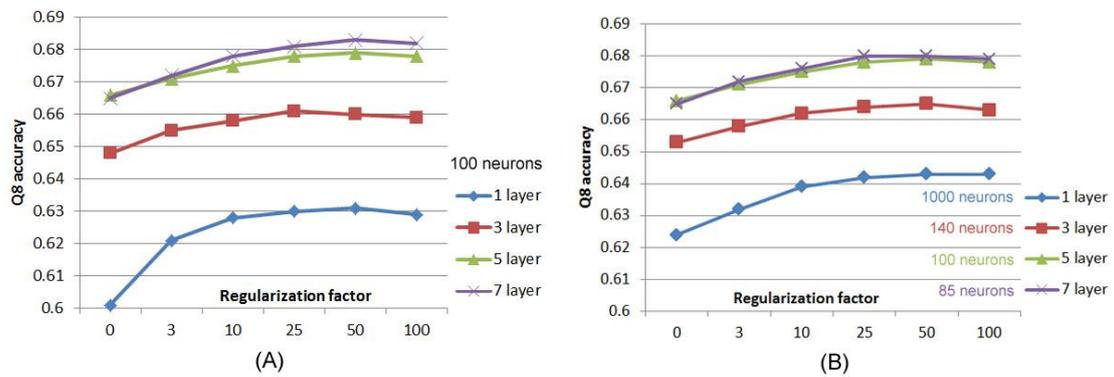

**Figure 4.** The Q8 accuracy on CB513 by the models of different number of layers of 1, 3, 5, and 7 (the same window size is used). (A) Each layer of the 4 models has 100 neurons for a position. The total parameter number of the 4 models is different. (B) Each layer of the models has different neurons for a position. The total parameter number of the 4 models is similar.

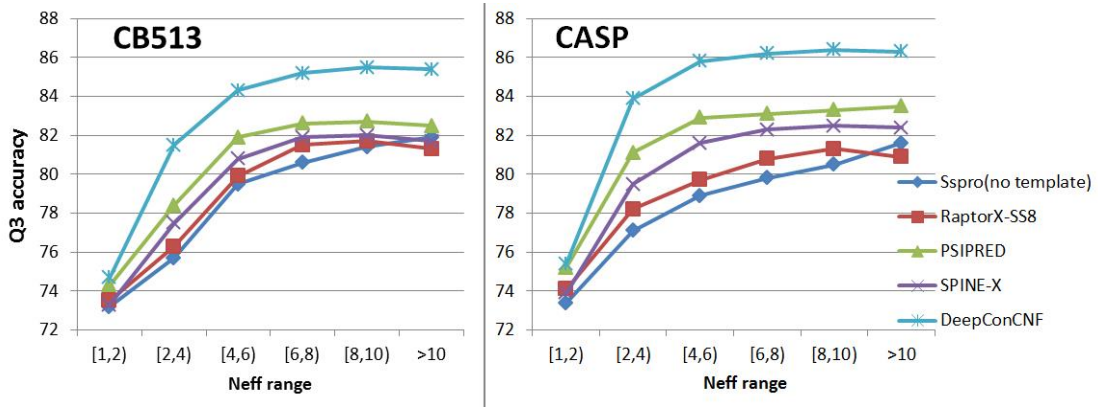

**Figure 5.** Q3 accuracy on CB513 and two CASP (CASP10-11) test sets with respect to Neff. Each point represents the average Q3 accuracy on those proteins falling into an Neff interval.

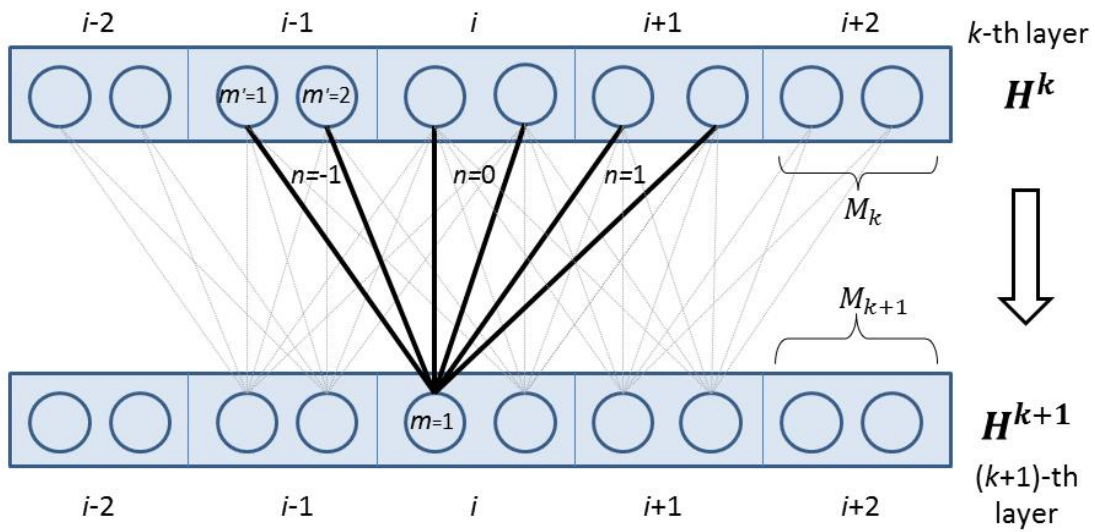

**Figure 6.** The feed-forward connection between two adjacent layers in the deep convolutional neural network.

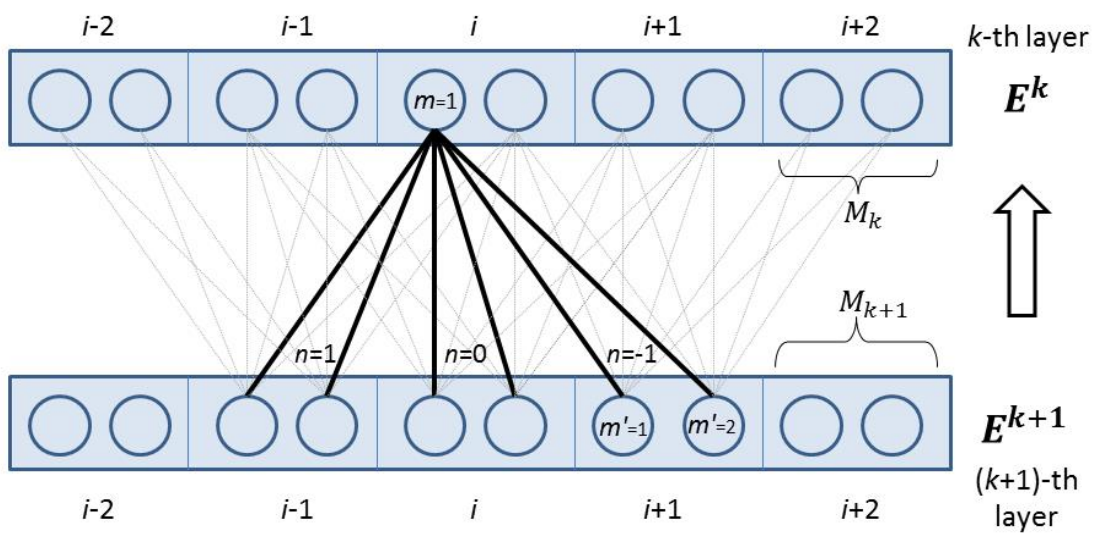

**Figure 7.** Illustration of calculating the gradient of deep convolutional neural network from layer $k+1$ to layer $k$.